\documentclass{article}
\usepackage{epsfig,graphicx}
\usepackage{float}
\usepackage{amsmath,amssymb}
\usepackage{authblk}
\newcommand{\be}{\begin{equation}}
\newcommand{\ee}{\end{equation}}

\newtheorem{theorem}{Theorem}[section]

\title{Epidemic model on a network: analysis and applications to COVID-19}

\author[1]{F. Bustamante-Casta\~neda\thanks{fbc.bercos.boson@gmail.com}}
\author[2]{J. G. Caputo \thanks{caputo@insa-rouen.fr}}
\author[3]{G. Cruz-Pacheco \thanks{cruz@mym.iimas.unam.mx}}
\author[2]{A. Knippel \thanks{arnaud.knippel@insa-rouen.fr}}
\author[4]{F. Mouatamide \thanks{fatza.mouatamide@gmail.com}}
\affil[1]{Posgrado de Matematicas, UNAM, Apdo. Postal 20--726, 01000 M\'{e}xico D.F., M\'{e}xico}
\affil[2]{Laboratoire de Math\'ematiques, INSA de Rouen Normandie\\ 76801 Saint-Etienne du Rouvray, France.}
\affil[3]{Depto. Matem\'{a}ticas y Mec\'{a}nica,
I.I.M.A.S.-U.N.A.M., Apdo. Postal 20--726, 01000 M\'{e}xico D.F., M\'{e}xico}
\affil[4]{University of Marrakech, Facult\'e des sciences Semlalia, Boulevard 
prince Moulay Abdellah, Marrakech 40000, Marocco. }

\date{\ }

\begin{document}
\maketitle
\vspace{-1cm}

\date{\ }

\begin{abstract}
We analyze an epidemic model on a network consisting of susceptible-infected-recovered 
equations at the nodes coupled by diffusion using a graph Laplacian.
We introduce an epidemic criterion
and examine different isolation strategies: we prove that
it is most effective to isolate a node of highest degree. The model
is also useful to evaluate deconfinement scenarios and prevent a so-called
second wave.  
The model has few parameters enabling fitting to the data and the essential
ingredient of importation of infected; these features are particularly
important for the current COVID-19 epidemic. 
\end{abstract}

\maketitle

\section{Introduction} 

Many models of the propagation of an epidemic such as the current
COVID-19 \cite{who} involve a network. This
can be a contact network between individuals. Then, the network
is oriented and is used to understand how a given individual
can infect others at the very early stages. The models 
are typically probabilistic, see
\cite{wsb17} for example. Once the epidemic is established, the
geographical network becomes important. There, nodes represent
locations and edges the means of communication; for COVID-19
these are the airline routes \cite{cbc20}. Such a network is non
oriented and the important nodes are the ones that are most connected.

One of simplest models of a disease is 
the Kermack-McKendrick system of equations \cite{km27} 
involving three populations of susceptible, infected
and recovered individuals $(S,I,R)$.
Using this model together with a probability transition matrix 
\cite{asmussen} for the
geographic coupling, Brockman and Helbling \cite{bh13} performed
a remarkable study of the propagation of well-known epidemics
like SARS or H1N1 due to airline travel. They emphasized that
the fluxes between the nodes govern the propagation of the epidemic.

In this article, we consider $(S,I,R)$ Kermack-McKendrick equations
coupled to a network through a graph Laplacian matrix \cite{crs01}. The
combination of the simple SIR dynamics with the diffusion yields the
essential ingredients to model and understand an epidemic, such as the
COVID-19. In particular,
\begin{itemize}
\item there are few parameters so that fitting to data can be successful,
\item it contains the essential ingredient of importation of infected subjects
from country to country.
\end{itemize}
The epidemic front is controlled by the availability of susceptibles.
If susceptibles are large enough, the front cannot be stopped. 
The number of susceptibles varies from node to node. 
Reducing this number at a given location can be done through
isolation. This is expensive and cannot be done for the whole network.
It is therefore important to address the question: 
what nodes are more useful to isolate to mitigate the epidemic?

Using this model together with the detailed data available \cite{jhopkins} \cite{world}, we 
predicted the onset of the COVID-19 epidemic in Mexico \cite{cbc20}. 
The present article is devoted to the detailed analysis of the model.
We first prove that it is well-posed and that
solutions remain positive. We introduce an epidemic criterion that
generalizes the well-known $R_0$ of the scalar case. For small
diffusion, nodes are almost decoupled and an outbreak
occurs at a node if the local $R_0$ is larger than one. When the diffusion
is moderate, the epidemic criterion depends on the network and 
when there is an outbreak, it starts synchronously on the network.
Using this criterion, we define an isolation policy. We find
that it is most useful to isolate the high connectivity nodes and not
efficient to isolate neighbors. For the particular case of the COVID-19
we discuss the effect of deconfining; the model shows that allowing 
circulation between heavily and weakly infected areas will prolong 
the outbreak in the latter.

The article is organized as follows. In section 2, we introduce the model,
discuss its main features and present the epidemic criterion. 
Section 3 shows a simple six node network based on the country of Mexico; 
there the effect of isolation is discussed. The COVID-19 disease is 
studied in section 4 and we show the
estimation of the time of outbreak in Mexico. The
important issue of deconfinement is studied in section 5. We conclude
in section 6.

\section{The model and epidemic criterion}

One of the main models to describe the time evolution of the outbreak 
of an epidemic is the Kermack-McKendrick model \cite{km27}
\begin{equation} \label{sir}
\left\{\begin{array}{ll}
  {\dot S}=- \beta SI, \\
  {\dot I} =\beta SI - \gamma I\\
  {\dot R} = \gamma I \\
 \end{array}\right .
\end{equation}
where the dynamics of transmission depends of the frequency and 
intensity of the interactions between (healthy) susceptible $S$
and infected individuals $I$ and produce recovered individuals $R$. 
The parameters $\beta$ and $\gamma$ are 
the infection rate and the recovery rate. 
The model conserves $N=S+I+R$ the total number of individuals.
Note that $R$ is essentially the integral of $I$
and therefore plays no role in the dynamics. We will omit it below and
only discuss $S$ and $I$.

An epidemic occurs if $\beta S - \gamma >0$ \cite{km27}. 
At $t=0$, $S=1$  so that an infection occurs if 
the infection factor defined as
\be\label{bigr}
R_0 \equiv {\beta \over \gamma}, \ee
is greater than one. 
An important moment in the time evolution of $S$ and $I$ is when
the number of infected is maximum. The corresponding
values $(S^*,I^*)$ can be calculated easily; we give the
derivation in the Appendix. The expressions are
\begin{eqnarray}
S^* = {1 \over R_0} , \\
I^* = I_0 +S_0- {1 \over R_0} (1+ \log (R_0 S_0) )
\label{sistar}
\end{eqnarray}
Note that $I^*$ and $S^*$ depend strongly on $R_0$. Take for example
$\gamma=0.625$ and different values of $\beta$.
\begin{center}
\begin{tabular}{|l|c|c|r|}
  \hline
$\beta$ & $R_0$  &  $S^*$  & $I^*$ \\ \hline
2.5      & 4      &  0.25   & 0.403   \\
1.5      & 2.4    &  0.417   & 0.218     \\
1.1      & 1.76   &  0.568   & 0.111     \\
  \hline
  \end{tabular}
  \end{center}
The value of $I^* $ depends also on the initial number of 
susceptibles $S_0$ which is smaller than the total number $N$.
Generally, a large $N$ gives a large $S_0$ and $I^* $.

\subsection{SIR on a network }

We consider a geographic network of cities connected by roads
or airline routes. This introduces a spatial component so 
that $(S,I)$ become vectors; we also drop $R$. This is similar to 
Murray's model where he introduces 
spatial dispersion in an SI model using a continuous Laplacian 
term \cite{murray}. 
The evolution at a node $j$ in a network of $n$ nodes reads
\begin{eqnarray} 
\label{dsij} 
{\dot S_j} = - \beta {S_j \over N_j} ~I_j  + \epsilon \sum_{k \sim j} (S_k-S_j), \\
{\dot I_j} = \beta {S_j\over N_j}  I_j - \gamma I_j + \epsilon \sum_{k\sim j} (I_k-I_j)  ,
\end{eqnarray}
where $N_j$ is the population at node $j$, the $\sum_{k\sim j}$ is the
exchange with the neighboring nodes $k$ of $j$ and where $\epsilon$
is a constant. The main difference with the model of \cite{bh13} is that
we assume symmetry in the exchanges.

The equations (\ref{dsij}) can be written concisely as
\begin{equation} \label{sil}
\left\{\begin{array}{ll}
  {\dot S} =\epsilon \Delta S - \beta S ~I, \\
  {\dot I} =\epsilon \Delta I  + \beta S I - \gamma I .\\
 \end{array}\right.
\end{equation}
where $S=(S_1,S_2,\dots,S_n)^T,~~ I=(I_1,I_2,\dots,I_n)^T,
~~\beta \equiv (\beta/N_1, \beta/N_2, \dots , \beta/N_n)^T$, 
$\Delta$ is the graph Laplacian 
matrix \cite{crs01} and we denote by $S I$ the vector \\
$(S_1 I_1 ,S_2 I_2,\dots,S_n I_n)^T$. 
{The infection rate $\beta$ can vary from one geographical site to another
while the recovery rate $\gamma$ depends only on the disease.
The diffusion $\epsilon$ should be small so that the populations
involved in that process remain much smaller than the node 
populations $N_j$. Another point is that the diffusion could act only on
the infected population. We chose to put the diffusion on both $S$ and $I$
for symmetry reasons. }

The graph Laplacian ${\Delta}$ is the real symmetric negative semi-definite 
matrix, defined as
\be\label{def_laplacian}
\Delta_{kl}= 1 ~{\rm if}~ k l ~{\rm connected}, ~0 ~{\rm otherwise} ; ~~ \Delta_{kk}= -\sum_{l \neq k}w_{kl}.\ee 
The graph Laplacian has important properties, see ref. \cite{crs01}, in
particular it is a finite difference approximation of the continuous
Laplacian \cite{ananum}. 
The eigenvalues of $\Delta$ are the $n$ non positive real numbers ordered and
denoted as follows:
\begin{eqnarray}
0=- \omega_1^2 \ge - \omega_2^2 \ge \dots \ge -\omega_n^2 .
\label{eigenvalues_Delta}
\end{eqnarray}
The eigenvectors $\{v^1,\dots,v^n\}$ satisfy 
\begin{equation}
\label{diagonalisation1}
{\Delta} {v}^{j}=-\omega _{j}^{2} {v}^{j}.
\end{equation}
and can be chosen to be orthonormal with respect to the scalar 
product in $\mathbb{R}^{n}$, i.e.
${v}^{i}\cdot {v}^{j} = \delta_{i,j}$ where $\delta_{i,j}$ is the Kronecker symbol.

\subsection {Well posedness and positivity }

The model (\ref{sil}) is well posed in the sense that the solution
remains bounded. We show this in the Appendix using
standard techniques.

The biological domain of the system is
$$\Omega = \{ (S,I): S\geq 0; I\geq 0\} .$$
Let us show that $\Omega$ is an invariant set for (\ref{sil})
so that the model makes sense in biology. Consider the different
axes $S_j=0$ and $I_j=0, ~j=1,\dots n$. First assume $I_j=0, ~j=1,\dots n$, 
then equation (\ref{sil}) reduces to
$$ {\dot S }  = \epsilon \Delta S $$
which conserves the positivity of $S $.
Similarly when $S =0$, we get
$${\dot I}  = \epsilon \Delta I - \gamma I$$
and again the positivity of $I$ is preserved.

\subsection{Epidemic criterion}

Here we extend the 1D epidemic criterion of Kermack-McKendrick \cite{km27}
to our graph model. Initially, the vector $I$ will follow 
the second equation of (\ref{sil})
\begin{equation} \label{it}
{\dot I} =(\epsilon \Delta -\gamma) I + \beta S I .
\end{equation}

Equation (\ref{it}) describes the onset of the epidemic on the
network. It can be written
$${\dot I}  = M I$$
where $M$ is the symmetric matrix
\be\label{mata} M = \epsilon \Delta - \gamma {Id}_n +{\rm diag}(\beta S_1,\beta S_2,\dots,\beta S_n).\ee
The eigenvalues of $M$ $\sigma_1,\dots,\sigma_n$ are real. If one of them is positive, then
the solution $I(t)$ increases exponentially and the epidemic occurs.
We can then write \\
{\bf Epidemic criterion : } there is an onset of the epidemic if
one eigenvalue $\sigma_i$ of $M$ is positive. 

{
Two situations occur, depending whether the diffusion is small
or moderate.
For small diffusion, the contribution of $\Delta$ to $M$ can be
neglected. Then each node will develop independently from the others.
We will have outbreaks in some and not in others.
}

When the diffusion is moderate, the Laplacian contributes to $M$.
Since $M$ is symmetric the eigenvalues of $M$ remain in the same order as
the ones of $\Delta$. This is the interlacing property \cite{crs01}. Then 
$\sigma_1$ will tend to 0 for $\gamma,\beta \to 0$. 
Note also that since $\parallel S \parallel$ decreases with time, the estimate
given by the eigenvalues of $M$ indicates the size of the epidemic i.e. 
${\rm max}~\parallel I \parallel$.
Then, the eigenvector of $M$ for the eigenvalue $\sigma_1$ will be almost
constant and the epidemic will start synchronously on the network.

The analysis of the moderate diffusion case can be extended 
when $\beta$ is constant.
Expanding $I$ on an orthonormal basis of eigenvectors ($v^k$) of $\Delta$
\be \label{iv} I=\sum_{k=1}^{n}  \gamma_{k}v^{k},\ee we get
\be\label{gkt1} {\dot \gamma_{k}}=(-\omega_{k}^{2}-\gamma)\gamma_{k}+
 <\beta  S I | v^{k}> .\ee
Assume that the susceptible population is constant on the network.
Then ${\rm diag}(S_1,S_2,\dots,S_n)= S {\rm Id}_n $ so that equation
(\ref{gkt}) reduces to
\be\label{gktc} {\dot \gamma_{k}}=(-\omega_{k}^{2}-\gamma +\beta S )\gamma_{k}  .\ee
The epidemic starts if $-\gamma +\beta S >0$ which is a simple generalization
of the criterion in the scalar case. \\
When the population of susceptibles is inhomogeneous and $\beta$ is
homogeneous, equation (\ref{gkt1}) becomes
\be\label{gkt} {\dot \gamma_{k}}=(-\omega_{k}^{2}-\gamma)\gamma_{k}+
\beta \sum_{l=1}^n \gamma_l \left ( \sum_{j=1}^n S_j  v_j^l v_j^k \right ) .\ee
Then the eigenvectors and the geometry of the network play a role.

\section{A simple example}

\begin{figure}[H]
\centerline{ \epsfig{file=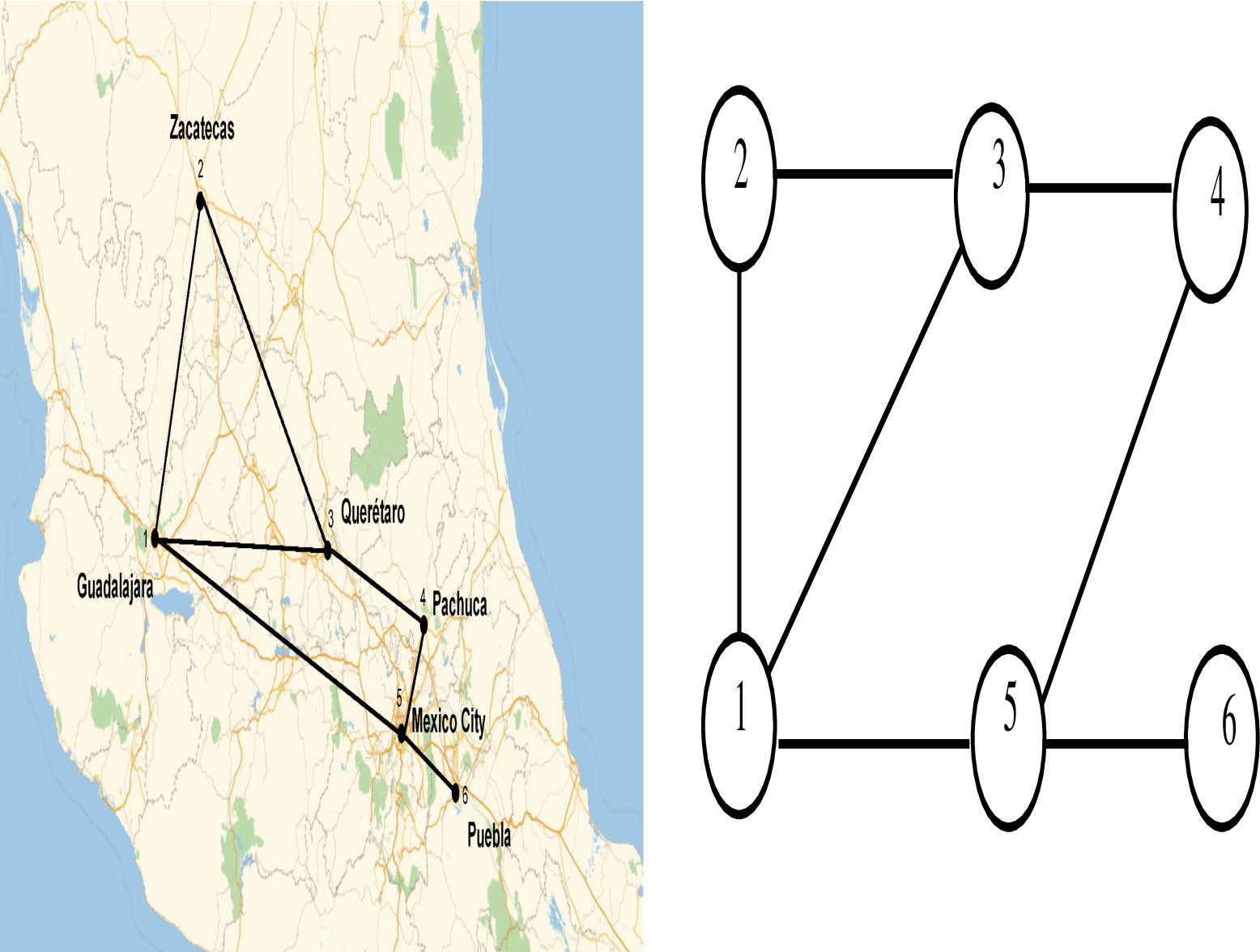,height=12 cm,width=4 cm,angle=90}}
\caption{
Graph of the six main cities in Mexico
numbered from 1 to 6: Guadalajara, Zacatecas, Queretaro, Pachuca,
Mexico City, Puebla. The links represent the main
roads connecting these cities. }
\label{g6}
\end{figure}

We illustrate the results given above on a 
6 node network inspired by the geographical map of
Mexico, with six main cities surrounding Mexico city, see Fig. \ref{g6}.
A node represents a city and an edge is a road link between two cities.
For simplicity, here we assume that $N_j$ is independant of $j$ so that
$I$ and $S$ are given in percentages. This will be the case throughout the
article unless specified.

The graph Laplacian is 
\begin{center}
$\Delta= \left[  \begin{array}{cccccc}
-3 & 1 & 1 &  0 & 1 &  0 \\
1  & -2&  1&  0 & 0 &  0 \\
1  &  1& -3& 1  & 0 &  0 \\
0  &  0&  1& -2 & 1 &  0 \\
1  &  0&  0&  1 & -3&  1 \\
0  &  0&  0&  0 & 1 & -1 
\end{array}\right] .$
\end{center}
The eigenvalues of this graph laplacian are
\begin{center}
\begin{tabular}{|c|c|c|c|c|c|}
  \hline 
0 & -0.721 & -1.682 & -3. &   -3.704 & -4.891 \\
  \hline 
  \end{tabular} 
  \end{center}
The corresponding eigenvectors are
\begin{center}
\begin{tabular}{|c|c|c|c|c|c|}
 \hline 
0.4082 & -0.2209 & -0.2007 & -0.5774 &  0.3084 &  0.5620 \\
0.4082 & -0.4149 & -0.5053 &  0.2887 & -0.5670 & -0.0323 \\
0.4082 & -0.3094 &  0.0403 &  0.2887 &  0.6581 & -0.4685 \\
0.4082 & -0.0692 &  0.7590 &  0.2887 & -0.2051 &  0.3564 \\
0.4082 &  0.2209 &  0.2007 & -0.5774 & -0.3084 & -0.5620 \\
0.4082 &  0.7935 & -0.2940 &  0.2887 &  0.1140 &  0.1444 \\
\hline 
\end{tabular} 
\end{center}

\subsection{Influence of the diffusion}

\begin{figure}[H]
\centerline{ \epsfig{file=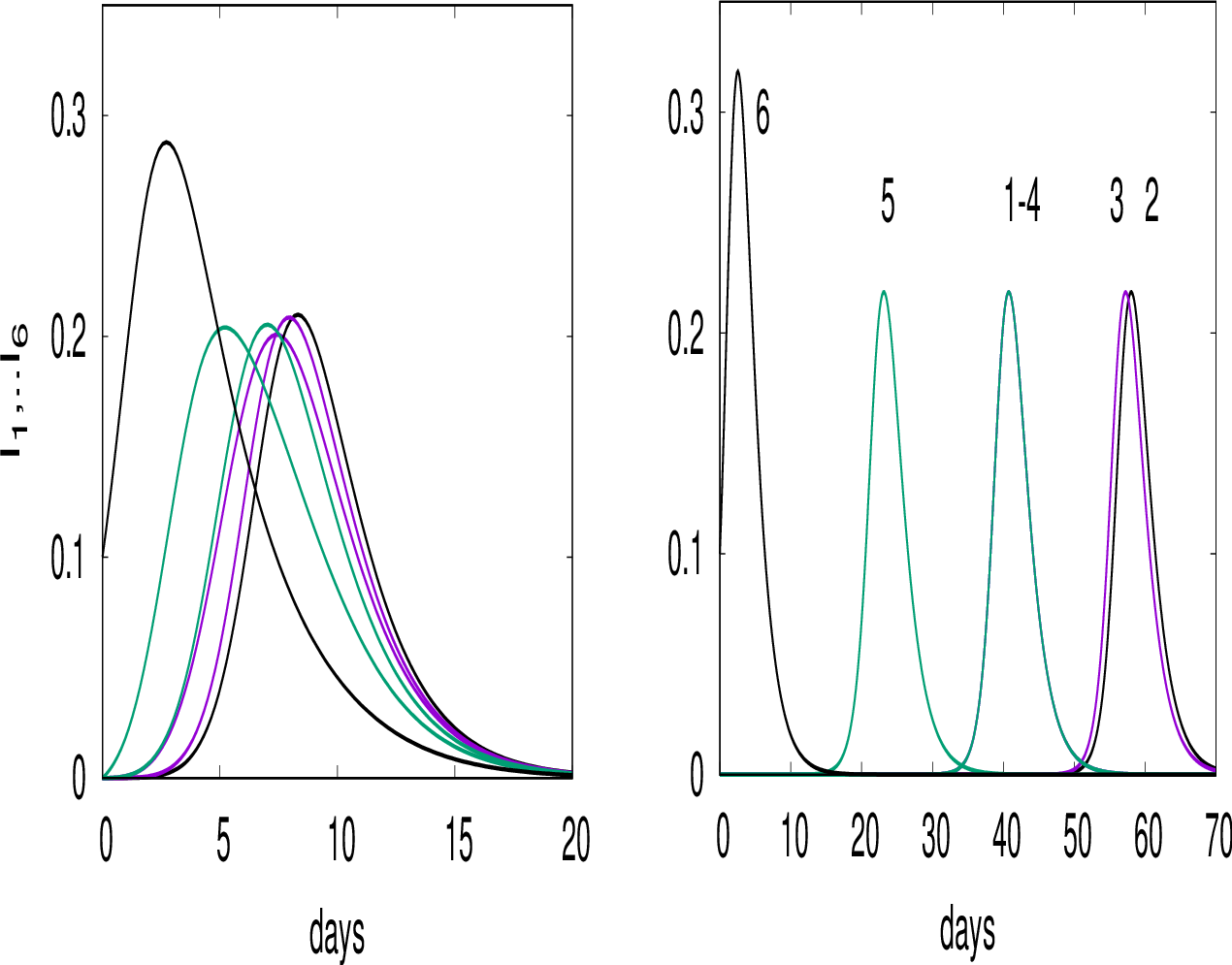,height=12cm,width=5cm,angle=90}}
\caption{Time evolution $I_k(t),~k=1,\dots,6$ for an outbreak at node 6
for $\epsilon =0.1$ (left panel) and $10^{-7}$ (right panel). The other parameters
are $\beta=1.5$ and $\gamma=0.625$.}
\label{m1}
\end{figure}

The variable $\epsilon$ measures the intensity of the diffusion
of $S$ and $I$ on the network. When $\epsilon << 1 $ the
diffusion is very weak and the evolution at each node can 
be decoupled from the one of its neighbors. 
For larger $\epsilon$, the diffusion and reaction occur on similar
time periods and need to be analyzed together. 
To see the influence of the diffusion, we plot in Fig. \ref{m1} 
the evolution of $I_k(t),~k=1,\dots,6$ for $\epsilon=0.1$ (left panel)
and $\epsilon=10^{-7}$ (right panel).
Note the times of arrival of the infection, first in 
node 5 the neighbor 
of node 6, then nodes 1 and 4 and finally nodes 3 and 2. For the
large diffusion (left panel of Fig. \ref{m1}) the peaks are very
close and there is a strong influence between the nodes.
On the other hand for a small diffusion, the peaks are well
separated and the nodes are decoupled. The maximum of $I_k$
is given by the estimate (\ref{sistar}).

\begin{figure}[H]
\centerline{ \epsfig{file=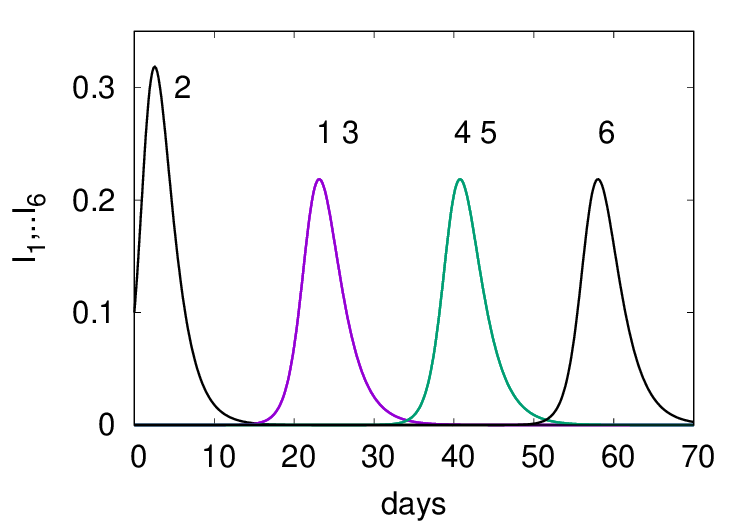,height=5cm,width=10cm,angle=0}}
\caption{Time evolution $I_k(t),~k=1,\dots,6$ for an outbreak at node 2.
The parameters are as in Fig. \ref{m1} (right panel).}
\label{m2}
\end{figure}

Infecting node 2 changes the time of arrival of the outbreak as
shown in Fig. \ref{m2}. It reaches first nodes 1 and 3, then nodes 4 and
5 and finally node 6.

\subsection{Isolation policies for large diffusion}

We first consider that the diffusion and the nonlinear terms
have similar orders of magnitude. We will address the case of weak
diffusion in the next section.

When the diffusion is large, one should consider the epidemic
on the network as a whole and use the topology of the network to
reduce the strength of the outbreak. 
From the amplitude equations (\ref{gkt}), one can devise a strategy
of isolation. By this we mean  reducing $\beta$ so that
the maximal eigenvalue of $M$ is minimum.

We choose 
$$\epsilon=1, ~~\beta=6.5, ~~\gamma=6.25 .$$
Table \ref{tab2} shows the eigenvalues $\sigma_1,\dots,\sigma_n$ of $M$ from (\ref{mata}) when
isolating a node of the network, i.e.
setting $S_j=0$ at a specific node $j$ and keeping the other nodes the same.
We chose $S=(1,1,1,1,1,1)^T$.
\begin{table} [H]
\centering
\begin{tabular}{|l|c|c|c|r|}
\hline
$j$ &  degree & $\sigma_3$ & $\sigma_2$ & $\sigma_1$  \\ \hline
6   &  1 & $-2.46$ & $-1.30$ & $ 1.22~10^{-1}$ \\
2   &  2 & $-2.14$ & $-7.85~10^{-1}$ & $ 3.83~10^{-2}$ \\
4   &  2 & $-2.51$ & $-4.79~10^{-1}$ & $ -1.06~10^{-2}$ \\
3   &  3 & $-1.44$ & $-8.29~10^{-1}$ & $ -3.71~10^{-2}$ \\
1   &  3 & $-1.52$ & $-6.41~10^{-1}$ & $ -6.72~10^{-2}$ \\
5   &  3 & $-1.52$ & $-6.41~10^{-1}$ & $ -6.72~10^{-2}$ \\
\hline
\end{tabular}
\caption{Isolated node $j$ and associated eigenvalues of $M$}
\label{tab2}
\end{table}
In the absence of isolation, the largest eigenvalue of the matrix
$M$ is $2.5 ~10^{-1}$.
The table shows that it is most effective to isolate nodes 1,3 and
5. These nodes have the highest degree of the network.

We now isolate two cities in the network. The results are presented
in table \ref{tab3}. 
We chose 
$$\epsilon=1, ~~\beta=6.75, ~~\gamma=6.25 ,$$
and $S= (1,1,1,1,1,1)^T$.
\begin{table} [H]
\centering
\begin{tabular}{|l|c|c|c|c|c|c|r|}
\hline
$i$ &  $j$ &  $d_i$ & $d_j$  & neighbors? &  $\sigma_3$ & $\sigma_2$ & $\sigma_1$  \\ \hline
 5& 6 &  3 & 1 & yes & $-2.97$& $ -1.27$ & $ 1.67~10^{-1} $ \\
 2& 3 &  2 & 3 & yes & $-1.99$& $ -9.49~10^{-1}$ & $ 1.56~10^{-1} $ \\
1& 2 &  3 & 2 & yes & $-2.57$& $ -6.83~10^{-1}$ & $ 1.27~10^{-1} $ \\
 4& 6 &  2 & 1 & no  & $-3.00$& $ -1.83$ & $ 9.17~10^{-2} $ \\
 1& 3 &  3 & 3 & yes & $-1.31$& $ -9.66~10^{-1}$ & $ 7.53~10^{-2} $ \\
 3& 4 &  3 & 2 & yes & $-2.26$& $ -6.24~10^{-1}$ & $ 6.01~10^{-2} $ \\
 2& 4 &  2 & 2 & no  & $-2.68$& $ -9.71~10^{-1}$ & $ 4.77~10^{-2} $ \\
 4& 5 &  2 & 3 & yes & $-2.82$& $ -3.94~10^{-1}$ & $ -9.08~10^{-3} $ \\
 1& 4 &  3 & 2 & no  & $-2.61$& $ -6.16~10^{-1}$ & $ -3.48~10^{-2} $ \\
 2& 6 &  2 & 1 & no  & $-2.33$& $ -1.89$ & $ -3.84~10^{-2} $ \\
 1& 6 &  3 & 1 & no  & $-2.60$& $ -1.14$ & $ -1.09~10^{-1} $ \\
 3& 6 &  3 & 1 & no  & $-2.41$& $ -1.06$ & $ -2.12~10^{-1} $ \\
 1& 5 &  3 & 3 & yes & $-1.39$& $ -4.24~10^{-1}$ & $ -2.50~10^{-1} $ \\
 2& 5 &  2 & 3 & no  & $-1.89$& $ -5.41~10^{-1}$ & $ -2.91~10^{-1} $ \\
 3& 5 &  3 & 3 & no  & $-1.33$& $ -5.97~10^{-1}$ & $ -3.37~10^{-1} $ \\
\hline
\end{tabular}
\caption{Isolated nodes $i,j$ and associated eigenvalues of $M$.}
\label{tab3a}
\end{table}
Again the high degree nodes 1,3 and 5 are the ones that reduce $\sigma_1$
the most and are therefore the most effective when applying isolation.
It is also not effective to isolate neighboring nodes.

The results shown in tables \ref{tab2} and \ref{tab3} 
can be explained in part from the properties of the matrix $M$
and the graph Laplacian $\Delta$. The maximal eigenvalue
$\sigma_1$ of $M$ verifies \cite{crs01}
\be\label{rayleigh}
\sigma_1= \sup_{\parallel X \parallel =1} X^T M X .\ee
We can find inequalities for $\sigma_1$ by choosing 
$$ X=(1,0\dots 0)^T,~X=(0,1,0\dots 0)^T,\dots$$
Denoting $d_i$ the degree of node $i$, we get
\begin{align}
\sigma_1 & \ge -\epsilon d_1 + \beta S_1 -\gamma,\\
\sigma_1 & \ge -\epsilon d_2 + \beta S_2 -\gamma,\\
\dots  \\
\sigma_1 & \ge -\epsilon d_n + \beta S_n -\gamma ,
\end{align}
so that
\be\label{vacsig1}
\sigma_1 \ge -{\rm min}_k d_k + \beta S -\gamma .\ee
This relation shows that isolating a node that has not smallest degree
does not change the estimate. Conversely, if there is a unique node of 
minimal degree and we isolate it, then the bound changes. \\
Using similar arguments, it can be shown that isolating 
two neighboring nodes, say 1 and 2 will be less effective
than isolating two non neighboring nodes.

Now we look at what happens if we cut a link, which corresponds to
condemning a road for example. Let $\Delta'$ be the Laplacian
of the new graph obtained by deleting a link. Without loss of generality
we can assume this link to be between vertices 1 and 2.
Then $\Delta'=\Delta -M$ where 
$$M = \begin{pmatrix} 
                 -1 & 1  & 0 & \dots & 0 \cr
                 1  & -1& 0 & \dots & 0 \cr
                 0 & \dots & \dots & 0
\end{pmatrix} $$
$M$ has all eigenvalues equal to $0$ except one which has value $-2$.
Applying the Courant-Weyl inequalities, see for example \cite{crs01}, we
get the following result for the maximum eigenvalue of $\Delta'$
$$\sigma'_1 \le \sigma_1.$$
Note that equality is possible: when $S$ is homogeneous, the maximum
eigenvalue of $M$ will always be $-\gamma + \beta S$. In such a case,
cutting a link is ineffective.

\subsection{Small diffusion and isolation policies}

\begin{figure}[H]
\centerline{ \epsfig{file=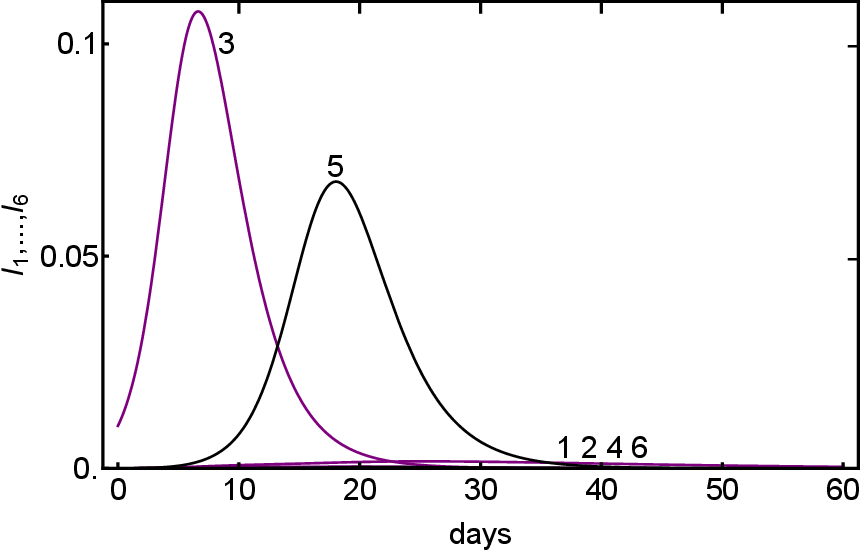,height=6 cm,width=10 cm,angle=0}}
\caption{Time evolution of the infected for different initial susceptibles
at the nodes. The parameters are $\beta= 1.98,~~ \gamma= 0.5 , ~~\epsilon= 10^{-3}$ and the initial conditions are given in table \ref{tab3}. }
\label{uneqs}
\end{figure}

When diffusion is small, the nodes evolve almost independently
so that the simple dynamics of the scalar SIR model apply. Then
some nodes can be isolated and the epidemic is not seen
there.  Fig. \ref{uneqs} shows such a situation. We choose
$$\beta= 1.98,~~ \gamma= 0.5 , ~~\epsilon= 10^{-3}$$ and the initial conditions
are given in the table \ref{tab3} below. The local $R_0$ is also
computed and one sees that an outbreak will occur at nodes
1,3 and 5 and not at nodes 2,4 and 6. Fig. \ref{uneqs} shows
the peaks for $I_3$ and $I_5$ and the maxima of $I_3$ and $I_5$ are
close to the ones predicted by the SIR formulas (\ref{sistar}).

\begin{table} [H]
\centering
\begin{tabular}{|l|c|c|c|c|c|r|}
  \hline
node $j$   & 1  &  2  &  3  &  4  & 5  & 6  \\ \hline
$S_j(t=0)$ & 0.26  &  0.14  &  0.55  &  0.16  & 0.5  & 0.18  \\
$I_j(t=0)$ & 0  &  0  &  0.01  &  0  & 0  & 0  \\
$R_0= \beta {{S_j}_0 \over \gamma} $ & 1.0296 &  0.5544 &  2.1780 &  0.6336 &  1.9800 &  0.7128 \\ 
$I^*$      & 0.0001 &  0.0364 &  0.1109  & 0.0227 &  0.0750 &  0.0130  \\
  \hline
\end{tabular}
\caption{Initial conditions and local $R_0$ for the plots of Fig. \ref{uneqs}.}
\label{tab3}
\end{table}

\section{ Propagation of COVID-19 }

\begin{figure}[H]
\centerline{ \epsfig{file=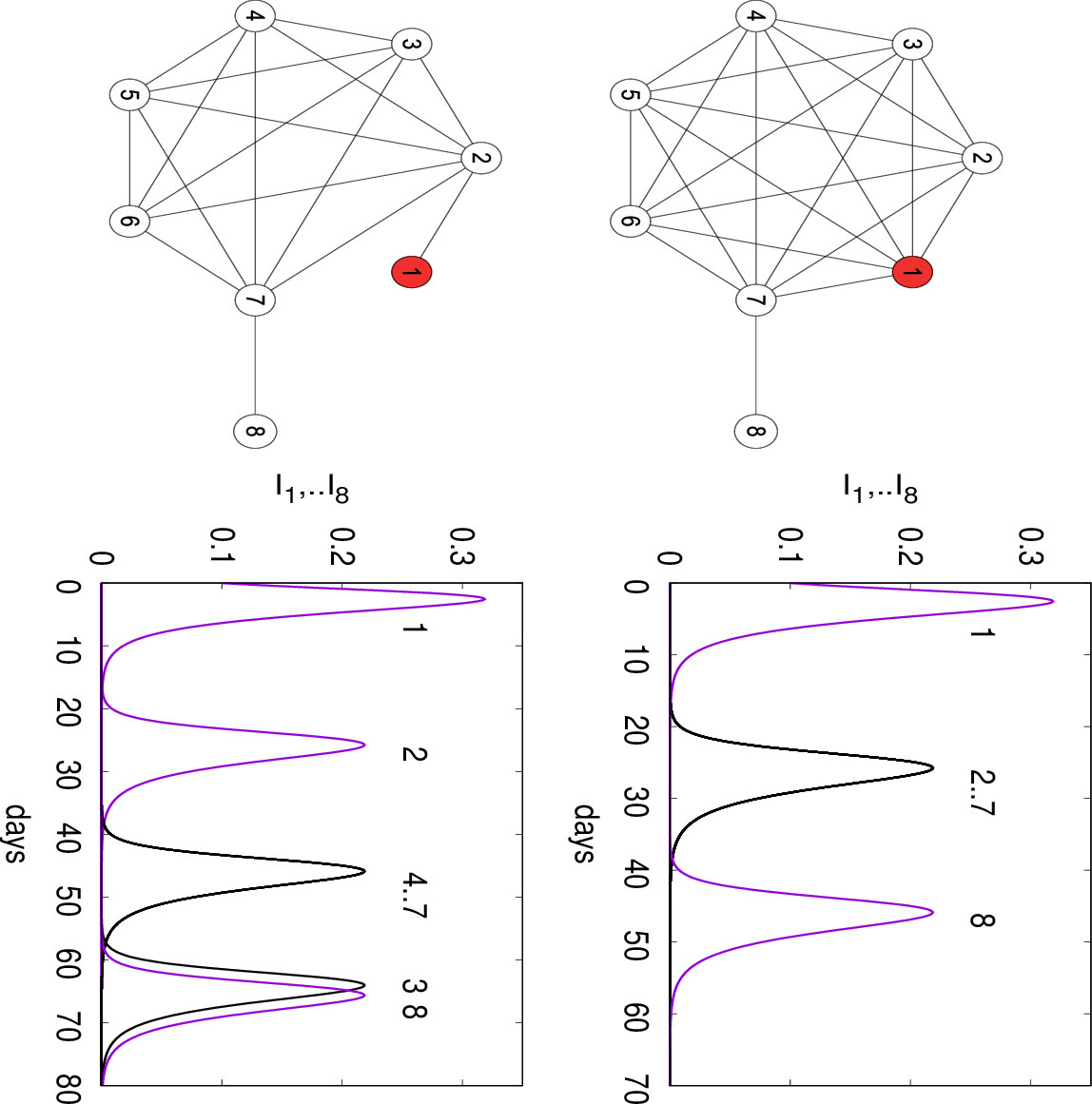,height=12cm,width=8cm,angle=90}}
\caption{ Propagation of the COVID-19: the network composed of seven
main regions/cities in the world 
forming a complete graph is shown in the top left panel and the 
time evolution of the infected per node is shown in the top right panel. 
In the bottom panels,
node 1 (Hubei) is only connected to node 2 (Beijing) and isolated from 
the other nodes:  see graph in 
the bottom left panel. The 
corresponding time evolution of the infected is shown in the bottom right
panel.
The parameters are  $\beta=0.5, \gamma=0.2$ and
$\epsilon = 10^{-6} {\rm days}^{-2}$. At time $t=0$ we
assume $I_1=0.1,~~I_k=0,~~k \neq 1$. }
\label{graph8}
\end{figure}

We now consider the propagation of the Corona virus COVID-19 on
a network consisting of a complete graph of 7 nodes with
an additional link to an 8th node, see top left panel of Fig. \ref{graph8}. 
The 7 nodes correspond to
the following cities or regions, Hubei, Beijing, Shanghai, Japan, 
western Europe, eastern USA and western USA and the 8th node is
Mexico. The links are the main airline routes. We assume a complete 
graph because airline routes connect any two of these regions.

The parameters chosen are  $\beta=0.5, \gamma=0.2$ and 
$\epsilon = 10^{-6} {\rm days}^{-2}$. These were suggested by very early
estimations of the outbreak in Wuhan. Using the data
from the John Hopkins website \cite{jhopkins} and our model,
we estimated the starting time of the outbreak in Mexico to be 
from March 20 to March 30 2020 \cite{cbc20}.
These results were shared with the Ministry of health of Mexico at the
end of February 2020 so that preparations could be made. 
In the present article, we analyze the model; we therefore comment
briefly the evolutions obtained in \cite{cbc20} as an interesting case.

The top right panel of Fig. \ref{graph8} shows the time evolution in days of 
the infected in the different nodes. The simulation is started 
at node 1 (Hubei in red in Fig. \ref{graph8}) with
$I_1=0.1$ and the other nodes are set at 0. The susceptibles are
set to 1 everywhere.
As expected the maximum $I_1^*=0.32$ and the subsequent maxima
$I_j^*=0.21,~~ j=2,\dots 8$. Notice how the nodes 2-7 start simultaneously
while node 8 is delayed. 
Communications with the province of Hubei were restricted at 
the end of December 2019.
To model this, we now consider that node 1 is only connected to node 2 which 
forms a complete graph with nodes 3-7, see bottom left panel
of Fig. \ref{graph8}. The infected are shown in the bottom right
panel; as expected the epidemic first arrives in node 2
then synchronously in nodes 3-7 and then in node 8.

To test the effect of having different $\beta$ s at each node
we increased $\beta_3$, reduced $\beta_4$ and kept the other $\beta$ s
the same. Then the epidemic arrives sooner at node 3 and later at node 4.
As expected from the formulas (\ref{sistar}) $I_3^* > I_j^* > I_4^*$
for $j=5,6,7$, see \cite{cbc20}.

\section{Confinement and deconfinement}
Since there is no vaccine for COVID-19 disease and the mortality
is relatively high, many countries put in place a confinement or
measures to reduce the movement of the population. 
\begin{figure}[H]
\centering
\centerline{ \epsfig{file=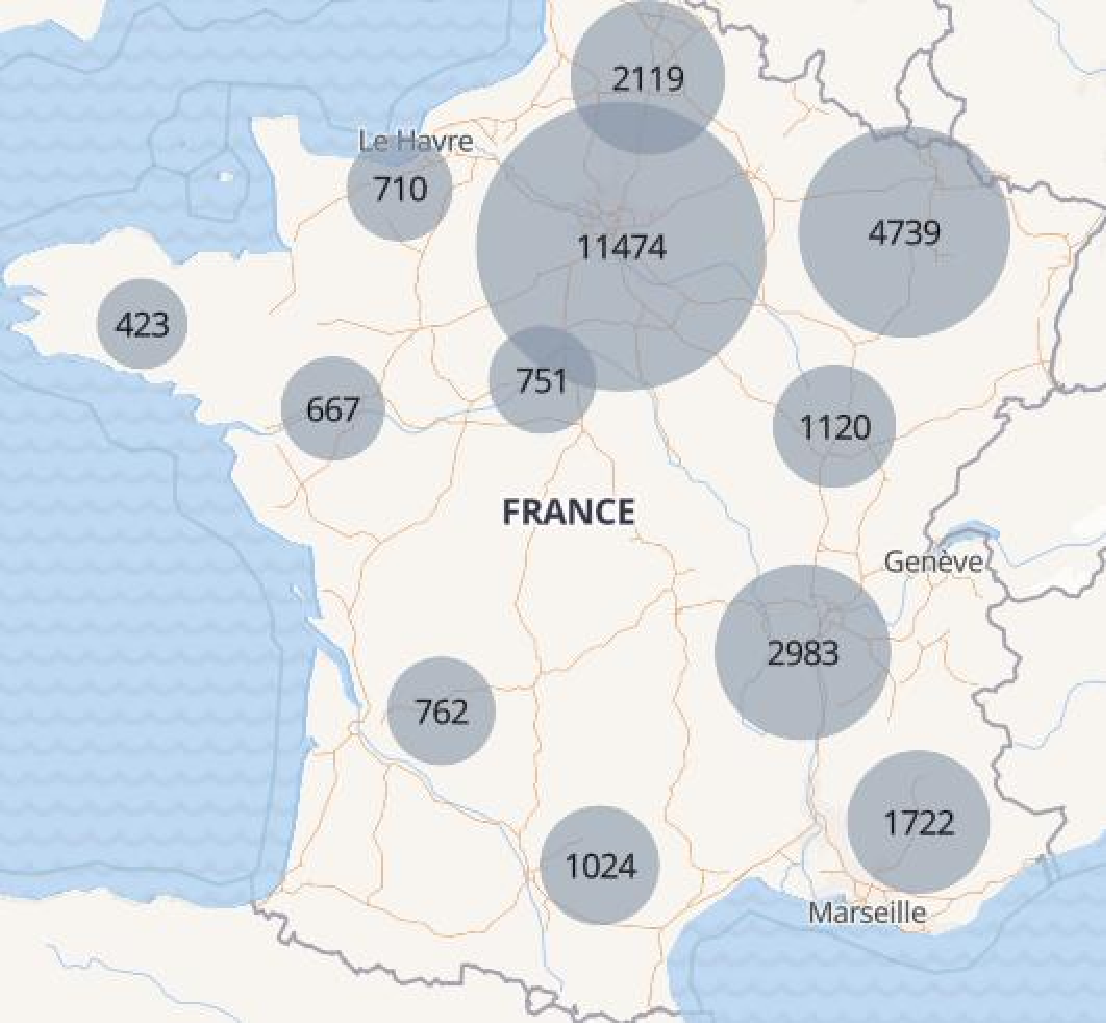,height=5 cm,width=10 cm,angle=0}}
\caption{Map of the number of patients in hospital due to COVID-19 in France
for the twelve different regions of France (see table \ref{tab4}) on April 5 2020.}
\label{fr5avril}
\end{figure}
{\small
\begin{table} [H]
\centering
\begin{tabular}{|l|c|r|}
\hline
region &  main cities & number of patients \\ \hline
Ile de France   & Paris &  11474  \\
Grand est       & Strasbourg, Mulhouse &  4739  \\
Auvergne-Rhone-Alpes     & Lyon, Grenoble, Clermont-Ferrand  &  2983  \\
Provence-Alpes-Cote d'Azur& Marseille, Nice & 1722  \\
Hauts de France &  Lille, Valenciennes & 2119 \\
Bourgogne-Franche Comt\'e & Dijon, Besancon & 1120  \\
Occitanie                 & Toulouse, Montpellier & 1024  \\
Nouvelle Acquitaine & Bordeaux & 762   \\
Centre Val de Loire       & Orleans, Tours & 751   \\
Normandie & Rouen, Caen, Le Havre & 710   \\
Pays de la Loire & Nantes & 667   \\
Bretagne & Rennes, Brest  & 423   \\ \hline
\end{tabular}
\caption{The twelve regions of mainland France, the main cities and 
the number of patients in hospital on April 5 2020.} 
\label{tab4}
\end{table}
}
China confined the
Hubei region around January 22, Italy confined its population
on March 9, France on March 17 and so on.
In the middle of the epidemic, Spain and France reached a situation 
like the one shown in Fig. \ref{fr5avril}. This picture
shows the number of patients in hospitals on April 5 2020 
for the 12 different regions
of mainland France (see Table \ref{tab4}), the data was obtained from the website
\cite{france}. Note how some regions are 
highly infected while others have many fewer cases.
To use this regional data of France, one could establish a graph
of the main roads and railways connecting the main cities, very much
like the one for Mexico in Fig. \ref{g6}.

Let us now consider the confinement. It can be implemented by 
\begin{itemize}
\item [(i)] reducing the contact ratio $\beta_j$ of each node $j$
\item [(ii)] reducing the diffusion $\epsilon$, ie the travel between nodes 
\end{itemize}
When deconfining the population once the peak
of the epidemic has passed the two options (i) and (ii) need to be relaxed,
so as to avoid a so-called second wave. This
happens in particular when the epidemic affected a small fraction of the
total number $N$. Then, relaxing $\beta$ or equivalently increasing $N$
causes a number of new susceptibles to enter the reaction and therefore
produce a second peak of infection. The model (\ref{dsij}) allows to analyze
the effect of the two options. We do this separately but note that in
reality both act together and effects can cancel.

\begin{figure}[H]
\centerline{ \epsfig{file=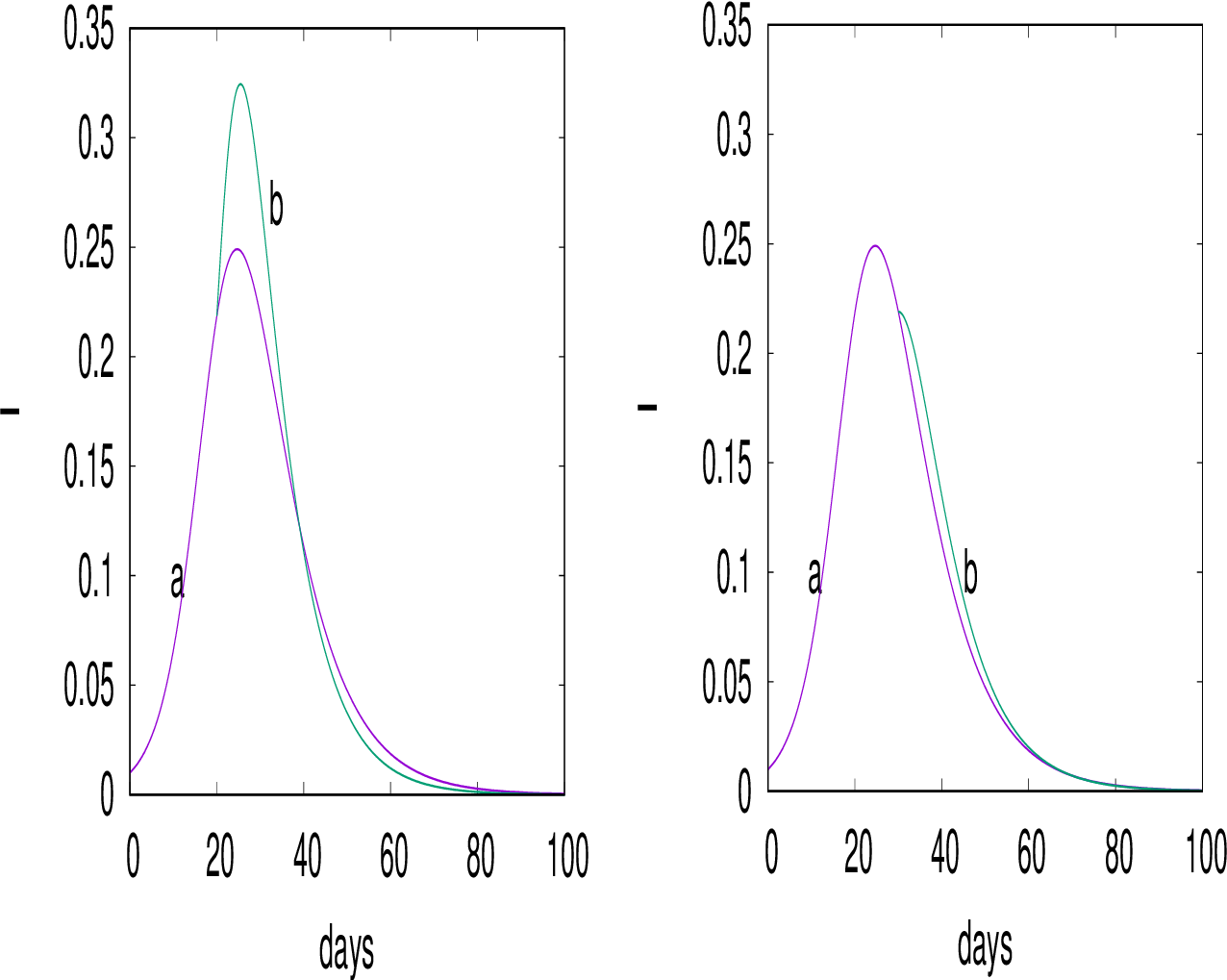,height=14cm,width=5cm,angle=90}}
\caption{\small\em Deconfinement (i) by increasing $\beta$ at a single node.
Evolution of $I(t)$ for a sudden increase of $\beta$ from 0.33 (curve a) to 
0.5 (curve b) at $t=20$
before the peak (left panel) and $t=30$ after the peak
(right panel); $\gamma=0.13$. 
}
\label{dec}
\end{figure}

Consider option (i). For this, we study the situation at a single node.
We choose the parameters
$$\beta=0.33,~~\gamma=0.13 , $$
and the computation is started at $t=0$ with $S_0=1,~I_0=0.01$.
Here, as before, the units of $I$ are percentages.
In Fig. \ref{dec}, we show the evolution of the infected for a 
sudden increase of $\beta$ from 0.33 to 0.5 at $t=20$
before the peak (left panel, line b ) and $t=30$ after the peak
(right panel, line b). Clearly, the deconfinement before the peak
causes many more infections and could saturate the hospitals. Deconfining
after the peak as shown on the right panel of Fig. \ref{dec} is harmless.
Only one node is involved. 

\begin{figure}[H]
\centerline{ \epsfig{file=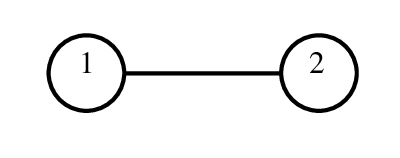,height=2 cm,width=4 cm,angle=0}}
\caption{The two node graph representing the interaction between a
large city and a small city.  }
\label{g2}
\end{figure}

Another way of deconfining is to relax option (ii), that is allowing travel from one node to another. This corresponds to increasing the diffusion which
can bring infected from large centers to small centers. To illustrate
this, consider the graph of two nodes
shown in Fig. \ref{g2}. It corresponds for the example of France 
Fig. \ref{fr5avril}, to allowing travel between the large urban area of Paris
and the much less populated Normandy. The capacity of node 1 (Paris) is
$N_1=20~ 10^6$ while $N_2= 10^6$. Now, $I_1$ and $I_2$ are
actual numbers and not percentages. We consider equations (\ref{dsij}) where
$S_j,I_j$ have been normalized by $10^6$. The parameters are
$$\beta_1= {\beta \over N_1}=0.025, ~~
\beta_2= {\beta \over N_2}=0.5, ~~\gamma=0.2, ~~\epsilon=10^{-6} .$$
The initial conditions are
$$S_1=20, ~~S_2=1,~~I_1=0.1,~~I_2=0.01 ,$$
in millions.
To model option (ii) we suddenly increase the diffusion parameter $\epsilon$
to $\epsilon=10^{-2}$.
\begin{figure}[H]
\centerline{ \epsfig{file=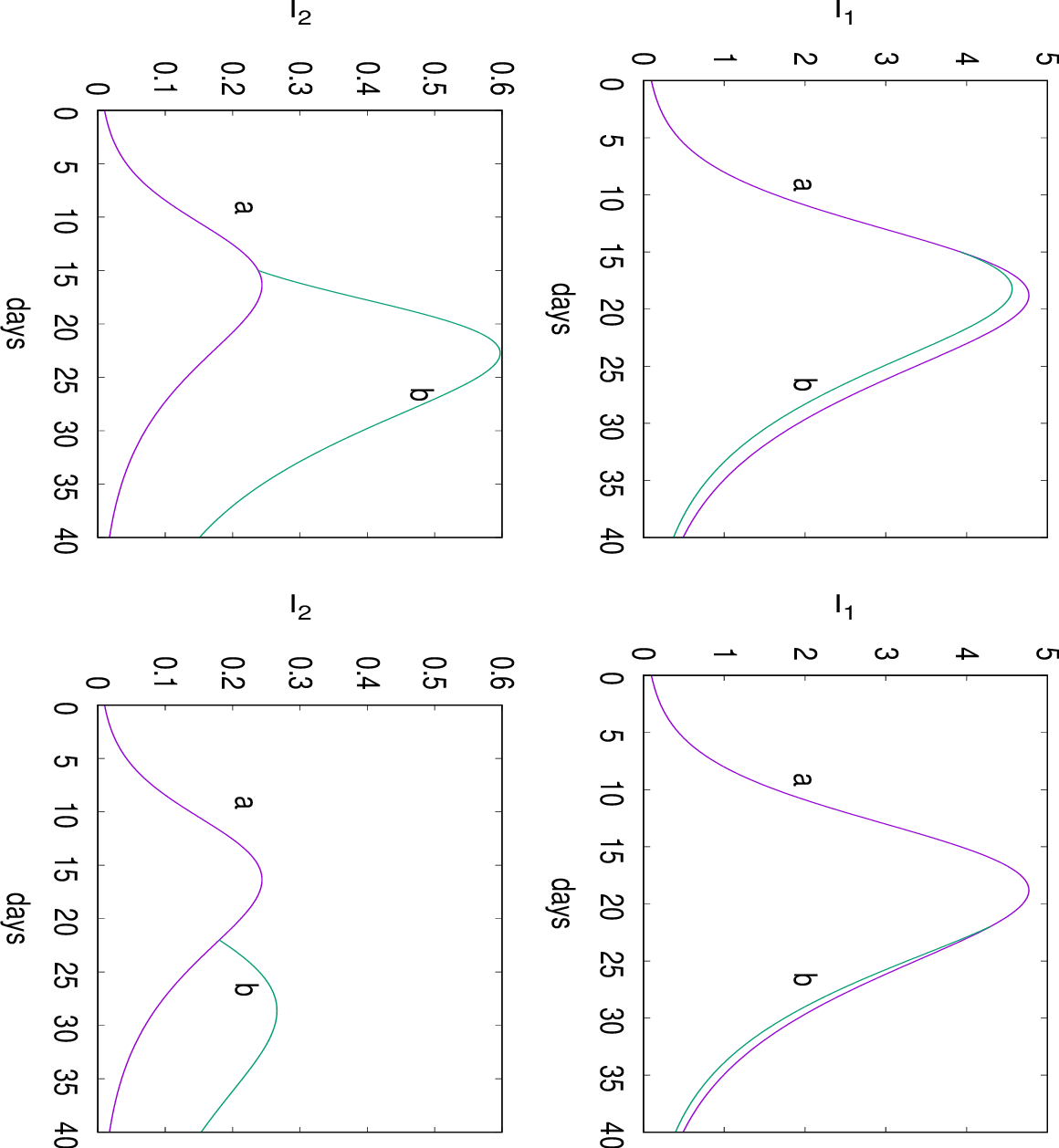,height=12cm,width=8cm,angle=90}}
\caption{\small\em Deconfinement (ii).  Time evolution $I_1(t),I_2(t)$ for the
two node graph shown in Fig. \ref{g2} when $\epsilon$ is
increased from $10^{-6}$ (curve a) to $10^{-2}$ (curve b), at $t=15$ 
before the epidemic peak
(left panel) and at $t=22$ (right panel) after the epidemic peak.
The units are in millions. See text for parameters and initial
conditions.
}
\label{conf}
\end{figure}
Fig. \ref{conf} shows $I_1(t)$ (top panels) and $I_2(t)$ 
(bottom panels) when deconfining before the peak (left panels)
and after the peak (right panels). The curves $I_1(t),I_2(t)$ 
before and after deconfining are indicated by $a$ and $b$ 
respectively. The $y$ scale are in millions.
Deconfining before the peak is not so harmful for node 1 while
it is catastrophic for node 2, $I_2$ is multiplied by 3. After the
peak, a sudden deconfinement prolongs the epidemic at node 2.

\section{Discusion and conclusion}

For the COVID-19 epidemic, few articles have addressed the coupling of
an epidemic model to the geographical landscape. The subject is difficult
as mobility is not well understood at microscopic level. It is also difficult
to extract a model from the available data.

In the reference \cite{china}, the authors estimated how 
mobility and transmissibility affect the onset of the epidemic in
the cities adjacent to Wuhan in the territory of China. 
They found that reducing the mobility between cities 
by some factor did not change
the time of onset of the epidemic peaks. This is
unexpected as seen in the present article where 
reducing the mobility by orders of magnitude delays the onset of the epidemic.
Also the world data \cite{jhopkins} and \cite{world} shows clearly the 
epidemic peaking first in China, then in Iran, Italy and so on.
Therefore, the result of \cite{china} might be due to 
the fact that small changes are effected on the mobility. It 
might also be due to the model of mobility chosen.

For Brazil, the study \cite{brasil} considers a six equation compartimental
model coupled to a complex mobility scheme. The authors show that the
epidemic curves vary "enormously" over different geographic scales.
Outbreaks can start in big cities and propagate to the countryside or there
might be multiple foci of infection. It is difficult to get a view of the
epidemic at the scale of the network with such a complex model.

In our study, we used the simplest susceptible-infected equations at the nodes
coupled by a geographic diffusion term. 
This contains the essential ingredient of importation of infected subjects
from country to country. 
We have kept the number of parameters to a minimum so 
that fitting the data can be successful. This is particularly
important for the present epidemic of COVID-19 where one wants to get 
a global picture at the level of the network.

The Laplacian models well the symmetric flow from one
node to another. This is a first order approximation of the spreading
of infected in terms of diffusion. One can refine the approximation
using a kernel of an anomalous or stochastic diffusion but then
more careful measures of the ways the infection travels are needed.

On the analysis side, using this model, we generalized the 
well-known epidemic criterion of Kermack-McKendrick.
For small diffusion, outbreaks occur at different times as the disease
advances through the network. A larger diffusion will cause the outbreak
to occur synchronously on the network. Using this criterion, we designed 
an isolation policy: we find it best to isolate 
high degree nodes and not efficient to isolate neighbors.

We also discussed the important aspect of deconfining a region after
the outbreak. Circulation between highly infected regions and less
impacted areas should be reduced 
to prevent the spread of infected to the latter. Large clusters should
be carefully controlled.

Finally the study points out the usefulness of having accurate data
at country and local levels (cities, neighborhoods and hospitals).

{\bf Acknowledgements} \\
This work is part of the XTerM project, co-financed by the European Union with the European regional development fund (ERDF) and by the Normandie Regional Council.

\appendix 
\section{Analysis of the SIR model}

The expressions 
(\ref{sistar}) may easily obtained in terms of $R_0$ as it is shown below.

From  $\dot{I}=0$ we obtain 
\be \label{sstar}
S^* = {\gamma \over \beta}= {1 \over R_0} . \ee
Dividing the second equation of (\ref{sir}) by the first, we get
$${d I \over dS} = -1 + {\gamma \over \beta S} ,$$
which can be integrated to yield
\be\label{iofs} I= I_0 +S_0-S +{\gamma \over \beta} \log {S \over S_0} , \ee
where we assumed $S(t=0)= S_0$ and $I(t=0)=I_0$.
Then one can compute $I^*$
\be \label{istar}
I^* = I_0 +S_0-S^* + {\gamma \over \beta} \log {S^* \over S_0}   . \ee
Assuming $S_0=1$, equations (\ref{sstar},\ref{istar}) can written in terms of $R_0$
as
\be
\label{sistar1}
S^* = {1 \over R_0},~~~~~
I^* = I_0 +S_0- {1 \over R_0} (1+ \log (R_0 S_0) )
\ee

The time $t^*$ corresponding to $S^*,I^*$ can be calculated in the
following way. \\
From the second equation of (\ref{sir}) one can write
$$ {dt \over dS} = -{1 \over \beta} {1 \over SI}= -{1 \over \beta} 
{1 \over S( I_0 + S_0 -S +{1 \over R_0} \log S)} ,$$
where we have substituted $I(S)$ from (\ref{iofs}).
Integrating this expression from $S^*$ to $S_0$ yields the value $t^*$ 
\be\label{tstar}
t^* = {1 \over \beta}\int_{S^*}^{S_0} 
{dS \over S (I_0 + S_0 -S +{1 \over R_0} \log S)} .\ee
This expression can be used to predict the time $t^*$ from
data.

\section{Well-posedness of the model }

To prove the well-posedness, we rewrite the system (\ref{sil}) 
as the following abstract differential equation:
\begin{equation}\left\{\begin{array}{ll}
x^{'}(t) = Ax(t) + f(x(t))\\
x(0)= x_{0}\in R^{n}\\
\end{array}\right. \label{Eq1}
\end{equation}
where $ x:=\left( \begin{array}{c}
s \\ 
i
\end{array} \right)  $, $ A $ is the matrix given by 
$$ A:=\left( \begin{array}{cc}
 \Delta & 0 \\ 
0 &  \Delta  
\end{array}  \right) $$ and $ f: R^{n}\times R^{n} \longrightarrow R^{2n} $ defined by $$ f(x):=\left( \begin{array}{c}
-\beta si \\ 
\beta si -\gamma i
\end{array} \right)  $$
and $ x_{0}:=\left( \begin{array}{c}
S_{0} \\ 
I_{0}
\end{array} \right)  $.\\

It is clear that, the function $ f $ is $L_{f}$-lipschitzian with $ L_{f} $ depends only on $ \beta$ and $\gamma$. Now, we formulate the well-posedness theorem, which is the main theorem  of this section:
\begin{theorem}
Given $ x_0 \in R^{n} $. Then, the equation \eqref{Eq1} has a unique solution satisfying the following formula:
\begin{eqnarray}
x(t)=e^{tA}x_{0}+ \int_{0}^{t} e^{(t-s)A}f(x(s))ds, \quad t\geq 0. \label{VarCons}
\end{eqnarray}
\end{theorem}

{\bf proof}\\

Let $x_0 \in R^{n} $ and $ T>0 $. Consider the mapping $ \Gamma: C\longrightarrow C $
given by $$ \Gamma u(t)= e^{tA}x_{0}+ \int_{0}^{t} e^{(t-s)A}f(u(s))ds $$
where $ C:= C([0,T], R^{n})$. Let us prove that $ \Gamma $ is a contraction. Indeed, let $ u,v \in C $, then 
\begin{eqnarray*}
\|\Gamma(u(t))-\Gamma(v(t)) \| &\leq & \int_{0}^{t}  e^{(t-s)\|A\|} \|f(u(s))-f(v(s)) \|ds \\ &\leq & L_{f}\int_{0}^{t}e^{(t-s)\|A\|}\| u(s)-v(s) \|ds \\ &\leq & L_{f} e^{T\|A\|}\int_{0}^{t}\| u(s)-v(s) \|ds \\  &\leq & L_{f} e^{T\|A\|}t \| u-v \|_{\infty}.
\end{eqnarray*}
On the other hand
\begin{eqnarray*}
\|\Gamma^{2}(u(t))-\Gamma^{2}(v(t)) \| &= & \|\Gamma(\Gamma u(t))-\Gamma(\Gamma v(t)) \|\\ &\leq & L_{f} e^{T\|A\|}\int_{0}^{t}s\| \Gamma(u(s))-\Gamma(v(s))\|ds \\  &\leq & \dfrac{(L_{f} e^{T\|A\|}t)^{2}}{2} \| u-v \|_{\infty}.
\end{eqnarray*} 
Hence, by iterating for $ n \geq 1 $, we conclude that 
\begin{eqnarray*}
\|\Gamma^{n}(u(t))-\Gamma^{n}(v(t)) \| &\leq & \dfrac{(L_{f} e^{T\|A\|}T)^{n}}{n!} \| u-v \|_{\infty}.
\end{eqnarray*} 
Now, for $ n $ large enough, $$ \dfrac{(L_{f} e^{T\|A\|}T)^{n}}{n!}<1 .$$ The mapping $ \Gamma^{n} $ is a contraction. Therefore, by using the iterating fixed point theorem $ \Gamma $ is also a contraction. Consequently, the system (13) has a unique solution which is given by (14).
{\bf end ~proof}

\end{document}